\documentclass[twocolumn,superscriptaddress,nofootinbib,floatfix,aps,prb,citeautoscript,longbibliography]{revtex4-2}
\usepackage[utf8]{inputenc}
\usepackage[T1]{fontenc}
\usepackage{lineno}

\usepackage{hyperref}
\usepackage{graphicx}
\usepackage{color}
\usepackage{array}
\usepackage{dcolumn}
\usepackage{bm}
\usepackage{multirow}
\usepackage[version=4]{mhchem}
\usepackage{cleveref}
\usepackage{tabularx}
\usepackage[usenames,dvipsnames]{xcolor} 
\usepackage{siunitx}
\usepackage{comment}

\newcommand{\Tc}{\ensuremath{T_\textrm{c}}}
\newcommand{\Tcmac}{\ensuremath{T_\textrm{c}^\text{McMillan}}}

\newcommand{\olog}{\ensuremath{\omega_\text{log}}}

\begin{document}

\newcommand{\halle}{Institut f\"ur Physik, Martin-Luther-Universit\"at
  Halle-Wittenberg, D-06099 Halle, Germany}
\newcommand{\coimbra}{CFisUC, Department of Physics, University of Coimbra, Rua Larga, 3004-516 Coimbra, Portugal}
\newcommand{\mpi}{Max-Planck-Institut f\"ur Mikrostrukturphysik, Weinberg 2, D-06120 Halle, Germany}

\author{Noah Hoffmann}
\affiliation{\halle} 
\author{Tiago F. T. Cerqueira}
\author{Pedro Borlido}
\affiliation{\coimbra}
\author{Antonio Sanna}
\affiliation{\mpi}
\author{Jonathan Schmidt}
\affiliation{\halle} 
\author{Miguel A. L. Marques} 
\email{miguel.marques@physik.uni-halle.de}
\affiliation{\halle} 

\date{\today}

\title{Searching for ductile superconducting Heusler X$_2$YZ compounds}

\begin{abstract}
Heusler compounds have always attracted a great deal of attention from researchers thanks to a wealth of interesting properties for technological applications. They are intermetallic ductile compounds, and some of them have been found to be superconducting.
With this in mind, we perform an extensive study of the superconducting and elastic properties of the cubic (full-)Heusler family. Starting from thermodynamically stable compounds, we use \textit{ab initio} methods for the calculation of the phonon spectra, electron-phonon couplings, superconducting critical temperatures and elastic tensors. By analyzing the statistical distributions of these properties and comparing them to anti-perovskites we recognize universal behaviors that should be common to all conventional superconductors while others turn out to be specific to the material family. The resulting data is used to train interpretable and predictive machine learning models, that are used to extend our knowledge of superconductivity in Heuslers and to provide an interpretation of our results.
In total, we discover a total of 8 hypothetical materials with critical temperatures above 10~K, to be compared with the current record of $\Tc=4.7$~K in this family. Furthermore, we expect most of these materials to be highly ductile, making them potential candidates for the manufacture of wires and tapes for superconducting magnets. 
\end{abstract}

\maketitle

\section{Introduction}

The image of a superconductor (likely an YBaCuO ceramic), immersed in liquid nitrogen and levitating over an array of magnets is undoubtedly familiar to anyone who has ever witnessed a science demonstration. These ceramics still hold the record for the highest superconducting transition temperature (\Tc) at ambient pressure (at around 133~K for \ce{HgBa2Ca2Cu3O_{1+x}}~\cite{10.1038/363056a0}), but other materials with high-\Tc\ have been found in the past decades~\cite{10.1088/1361-648x/ac2864}, e.g. \ce{MgB2} ($\Tc=39$~K~\cite{nagamatsu2001superconductivity}), fullerides such as \ce{Cs3C60} ($\Tc=38$~K~\cite{10.1038/nmat2179}), thin films of FeSe ($\Tc>100$~K~\cite{10.1038/nmat4153}), etc. More recently, hydrides with exceptionally high critical temperatures were also discovered, but at very high pressure~\cite{10.1016/j.physrep.2020.02.003}.

In spite of these remarkable advances, to this day, niobium containing materials discovered in the 1950's and 1960's are still the go-to choice for commercial applications~\cite{narlikarSuperconductors2014}, the most relevant of which are niobium-titanium (Nb--Ti) alloys. Notably, this happens in spite of their maximum critical temperature of 9.8~K at 24~wt\% Ti~\cite{berlincourtSuperconductivityHighMagnetic1963}, which pales in comparison with the previous examples.
\ce{NbSn3} is another commercial superconductor, presenting not only a higher critical temperature of 18.5~K, but more importantly a larger critical field of 30~T~\cite{dew-hughesSuperconductingA15Compounds1975}. Because of this, it finds use in applications requiring much larger operating magnetic fields than those attainable by Nb--Ti alloys. The prototypical example of this are the operating electromagnets of the International Thermonuclear Experimental Reactor (ITER), where Nb--Ti wirings are supplemented with \ce{Nb3Sn} inner windings.

Looking at metrics like critical fields and temperatures alone, it is hard to understand why Nb--Ti has not been entirely replaced by \ce{Nb3Sn} (nor by any other high-\Tc\ superconductor) as the industry standard. It is true that these two properties are necessary for a `good' superconductor, but they are not sufficient from an engineering point-of-view, as a more critical aspect is the ability to draw material into continuous wire or tape several kilometers long with consistent fabrication quality. This, generally speaking, translates into a need for ductile materials. For example, although \ce{Nb3Sn} is used in devices, manufacture of wires is complicated due to its brittleness, and requires complex production methods leading to higher fabrication costs~\cite{narlikarSuperconductors2014}. Higher-temperature superconductors, such as \ce{MgB2} or the ceramics are even more brittle than \ce{Nb3Sn}, leading to even more complex manufacturing problems.

Several requirements come to mind in the search for new superconductors that can replace Nb--Ti alloys in commercial applications -- ductility, lower density to accommodate easier transportation, higher critical field, no Nb which is considered a critical raw material by the European Union~\cite{euro_report}, and compatibility with available production methods. Broadly speaking, these conditions point towards intermetallic compounds (as the presence of non-metallic elements often leads to brittle materials) and with first- or second-row metallic elements (where superconductivity is usually driven by the conventional electron-phonon mechanism).
These systems can be treated in a straightforward manner by modern \textit{ab-initio} techniques: electron-phonon superconductivity is well understood~\cite{10.1088/1361-648x/ac2864}, and several electronic structure packages implementing some form of Eliashberg theory for the calculation of critical temperatures exist. Mechanical properties such as ductility, trivially treated at the macroscopic level, are harder to translate in terms of atomic calculations but nonetheless some models are available~\cite{senkovGeneralizationIntrinsicDuctiletobrittle2021}. Finally, machine-learning methods can be used to accelerate the calculations and to help in the interpretation of the results~\cite{10.1088/2516-1075/ac572f,10.1038/s41524-019-0221-0}.

Studying all intermetallic prototypes is at the moment untractable, and thus we will focus on a specific family of compounds, namely the (full-)Heuslers. Named after Fritz Heusler, these are a large class of intermetallic compounds which crystallize in a face-centered cubic structure and have \ce{X2YZ} composition, with X and Y are transition metals and Z is a main group metal~\cite{Bai2012_1230006}. Heuslers possess a wide range of compositions and tunable material properties, making them an ideal family to search for ductile superconductors. They have been researched in areas as diverse as thermal conductivity~\cite{carreteFindingUnprecedentedlyLowThermalConductivity2014}, thermoelectricity~\cite{carreteNanograinedHalfHeuslerSemiconductors2014, raghuvanshiHighThroughputSearch2020, sakuradaEffectTiSubstitution2005}, topological insulators~\cite{chadovTunableMultifunctionalTopological2010, linHalfHeuslerTernaryCompounds2010} and magnetism~\cite{kainumaMagneticfieldinducedShapeRecovery2006, krenkeInverseMagnetocaloricEffect2005, vanengenPtMnSbMaterialVery1983}.

The first superconducting Heusler compounds, found by Ishikawa et al. in 1982 \cite{ishikawa1982superconductivity}, were of the form \ce{Pd2REPb}, where \ce{RE} is a rare earth metal. Shortly after, in 1983, Wernick et al. discovered superconductivity in \ce{Ni}-based systems~\cite{wernick1983superconductivity}. Since then, several Heusler compounds were found to be superconducting~\cite{klimczukSuperconductivityHeuslerFamily2012,
  malikMagneticMossbauerStudies1985,
  pooleHandbookSuperconductivity2000, rameshkumarVanHoveScenario2013,
  seamanSuperconductivityMagnetismHeusler1996,
  sheltonCoexistenceSuperconductivityLongrange1986,
  suhlSuperconductivityFbandMetals1980,
  wakiSuperconductivityNiNbX1985, winiarskiMgPdSb2021,
  winterlikNibasedSuperconductorHeusler2008}. This even includes
compounds with a coexisting magnetic and superconducting state, e.g. \ce{Pd2YbSn}~\cite{kierstead1985coexistence} or \ce{Pd2ErSn}~\cite{sheltonCoexistenceSuperconductivityLongrange1986}. Up to now, the record \Tc\ belongs to \ce{Pd2YSn} with 4.7~K~\cite{wernick1983superconductivity}, followed by \ce{Au2ScAl}
with 4.4~K~\cite{pooleHandbookSuperconductivity2000}. Unfortunately the elastic properties of full-Heuslers, even though extensively studied in theory works, are much less explored experimentally (see Ref.~\cite{Everhart2019} and references therein), being somewhat easier to find information on the half-Heusler family~\cite{He2015, Rogl2016, Everhart2019}.

Here, we perform an extensive study of the superconducting and elastic properties of Heusler materials. These are then compared to a very different family of compounds, specifically the anti-perovskites, that some of us studied recently~\cite{oursperovskites}. This is interesting, as many anti-perovskites are also superconducting, but contain a non-metallic element (such as C, O, N, etc.). This comparison then allows us to discern between properties specific to the Heusler family from behavior generally present in all electron-phonon driven superconductors.

The remainder of the work is divided as follows. First we discuss the two steps of high-throughput calculations, (one for the full-Heuslers near the convex hull of stability, and another for the remaining full-Heuslers within 200 meV/atom of the hull), accompanied by a general discussion of  the distribution of relevant properties (e.g. $\lambda$, \olog, or \Tc). Due to the number of meta-stable materials, the second step was accelerated with the machine learning models trained with the results from the first step. For the materials with the best performing critical temperatures, mechanical properties were calculated in an attempt to find those most likely to be ductile. Lastly, we perform a detailed analysis of the materials that we considered to be the best overall. A flowchart showing these various steps is shown in Figure~S1 of the Supplementary Material (SM).

\section{Results and Discussion}

\subsection{High-throughput}

There are several high-throughput studies of the thermodynamic stability (and other properties) of Heusler compounds~\cite{10.1021/acs.chemmater.6b02724, 10.1016/j.jallcom.2021.158854}, and ground-state calculations for essentially all compounds of this family can be found in several databases~\cite{aflowlib,NOMAD,CGAT}.
Our present analysis begins with the dataset of Ref.~\onlinecite{CGAT} from which we selected all compounds that are metallic and that lie on (or very close to) the convex-hull of stability, as calculated with the PBE exchange-correlation functional. In order to avoid problems associated with magnetism and superconductivity, only materials with a non-magnetic ground-state were considered. This selection resulted in a total of 565 materials, with the full list given in the SM.

\begin{figure*}
  \centering
  \includegraphics[height=4.25cm]{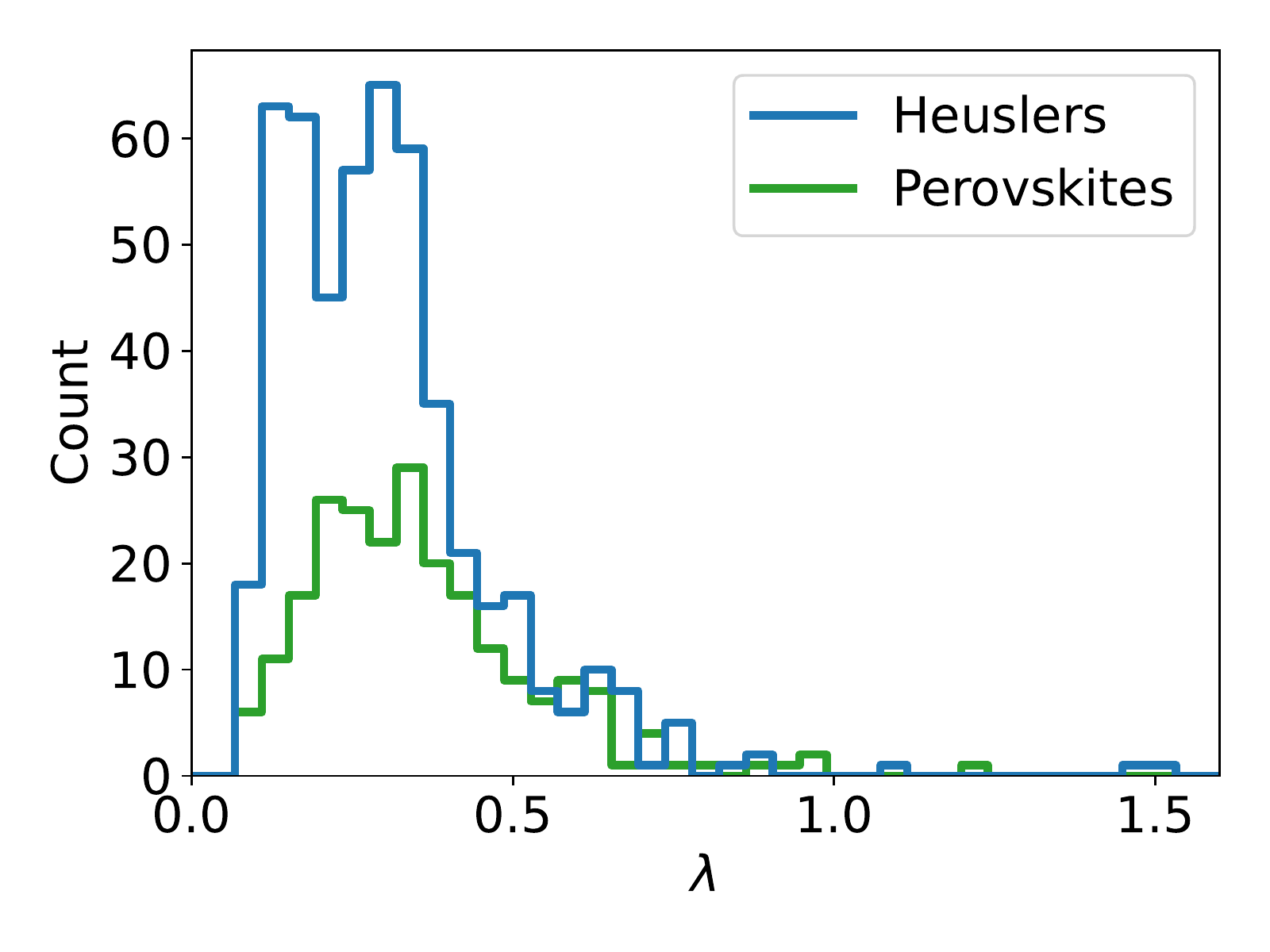}
  \includegraphics[height=4.25cm]{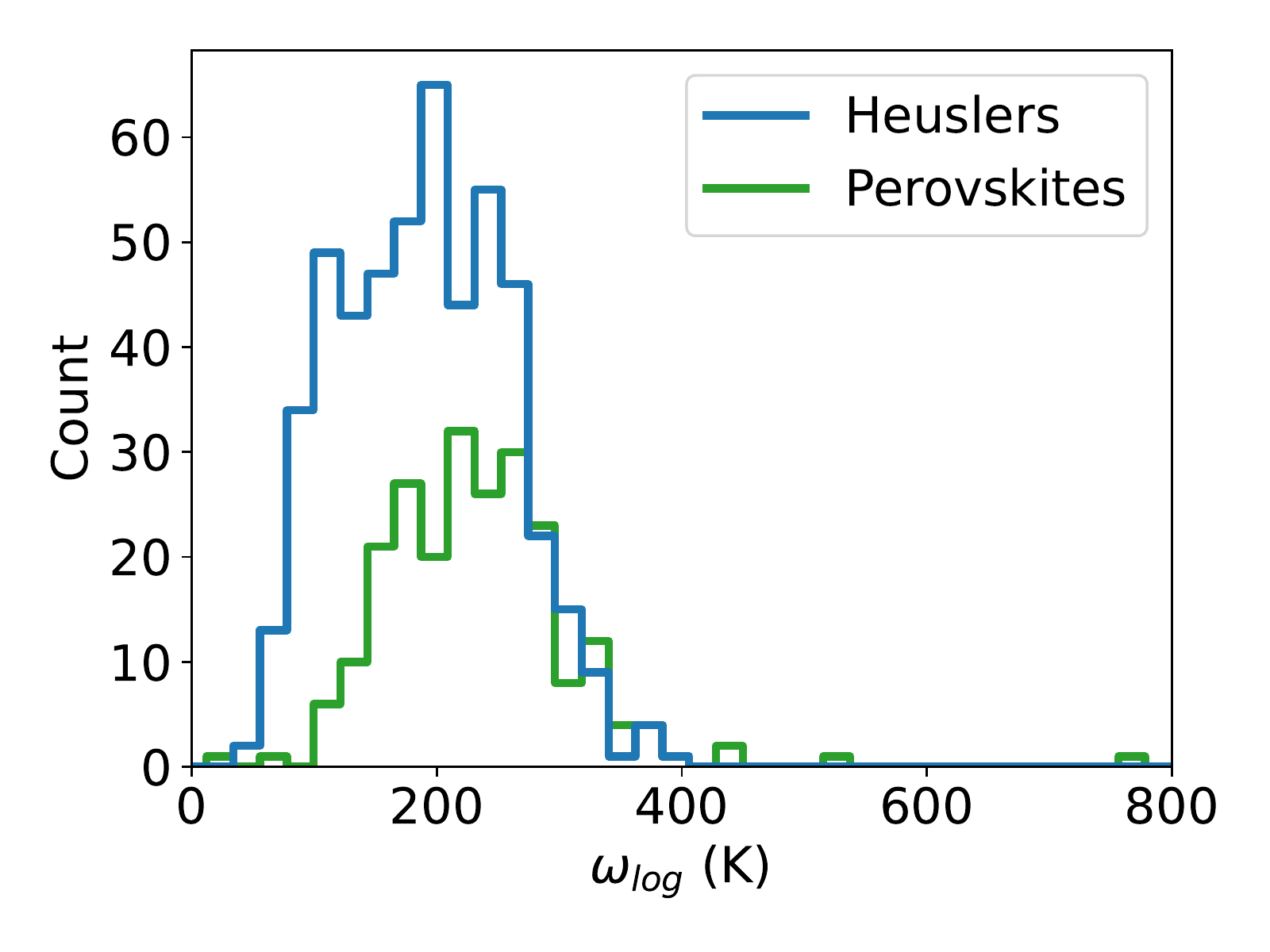}
  \includegraphics[height=4.25cm]{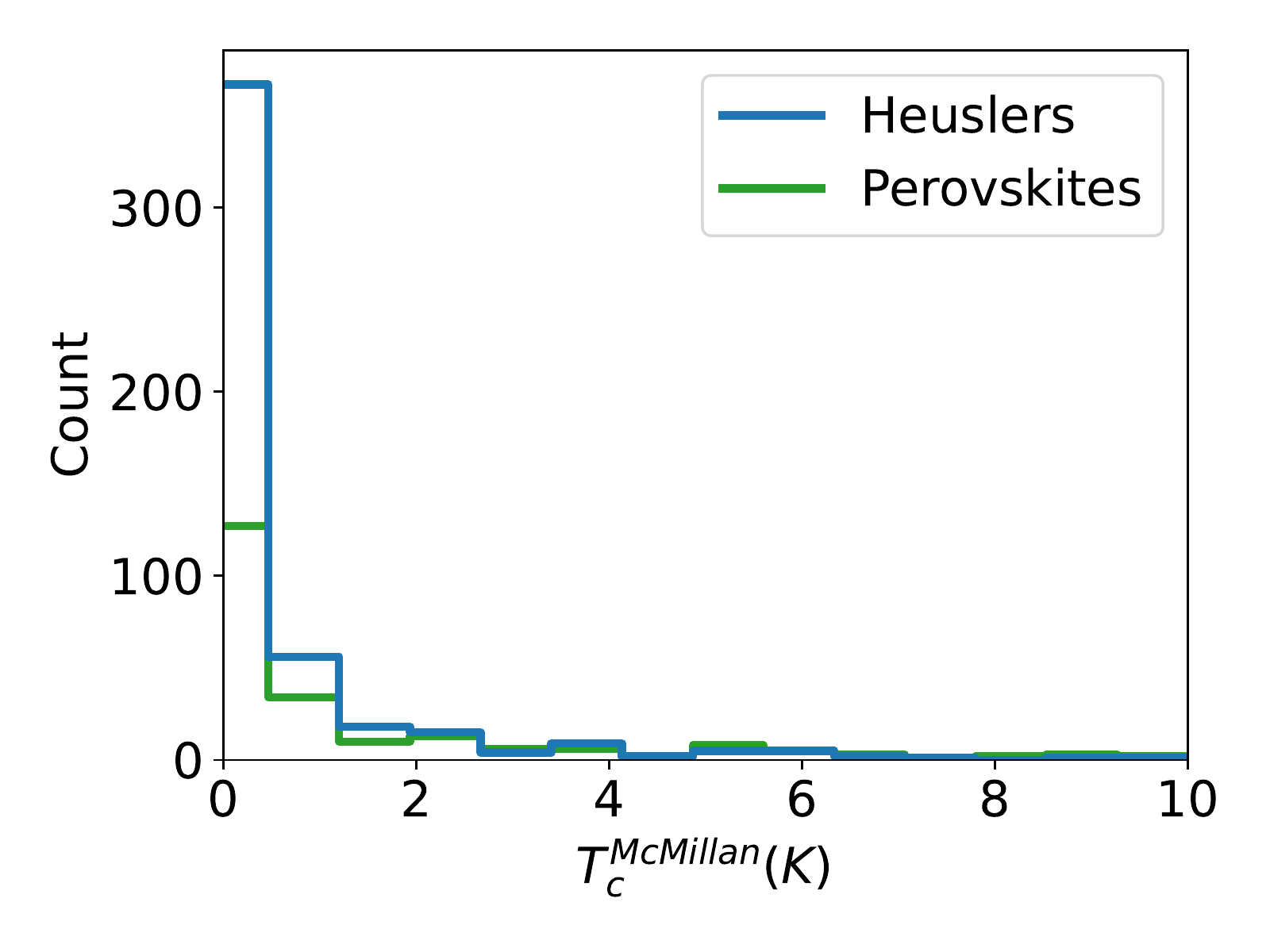}
  \caption{Histograms of the calculated values (with a $4\times4\times4$ $q$-point grid) of the
    electron-phonon mass enhancement parameter $\lambda$, the averaged phonon
    frequency $\omega_\text{log}$ (in K), and the superconducting transition
    temperature $T_\text{c}$ (in K; calculated with the McMillan formula). The blue curves are for Heuslers in our training set and the green curves are for the anti-perovskites contained in the training set of Ref.~\onlinecite{oursperovskites}.}
    \label{fig:lambda_omega_tc_histograms}
\end{figure*}

For these systems, the phonon dispersion curves were calculated which resulted in further removing several entries due to the presence of imaginary modes, resulting in 502 entries. Finally, for the remaining dynamically stable systems we calculated the electron-phonon mass enhancement parameter, $\lambda$, and the logarithm averaged phonon frequency, $\omega_\text{log}$. 
From these, the critical temperature, $T_\text{c}$, using McMillan's formula~\cite{Mcmillan1968tc}, as well as Allen–Dynes' modified formula~\cite{Allen1975}, were computed using a constant value of $\mu^*=0.1$ as detailed in \cref{sec:methods}. For materials with a McMillan temperature higher than 1~K we also computed the critical temperature using the isotropic Eliashberg equation~\cite{IsotropicEliashberg}. All these values can be found in the SM.

At this point we must notice that several structures that the harmonic approximation at 0~K and 0~GPa predicts to be dynamically unstable are known to be synthesizable at room conditions due to anharmonic and entropic effects. This occurs in several superconducting Heuslers as reported in Table~S1 of the SM. Discarding structures with imaginary frequencies may certainly lead to missing some superconductors but the computational cost associated with including higher-order effects makes the calculations prohibitive for an high throughput search. As such, at the present we ignore these effects and hope they can be addressed in future work.

With respect to the different approaches for calculating \Tc\, we see that both McMillan and Allen-Dynes underestimate the critical values with respect to the Eliashberg one, although these are nonetheless strongly correlated (see Figure~S2 of the SM). Given that the former two values are considerably simpler to obtain, they are a good quantity to use in high-throughput studies and as input for machine-learning studies. In the following, to avoid confusion and without loss of generality, we thus give preference to \Tcmac\ in the discussion of distributions and only refer to the other values in more specific cases.

Histograms for the resulting values of $\lambda$, $\olog$ and $\Tcmac$ obtained with the 4x4x4 set of parameters (see \cref{sec:methods}) are presented in Figure~\ref{fig:lambda_omega_tc_histograms}. 

The parameter $\lambda$ follows an asymmetric distribution, akin to a Poisson distribution but with a slower decaying tail for large values.
With a mean value of $0.30$, most Heuslers must be considered to have weak electron–phonon coupling. We found a few compounds with larger values of $\lambda > 1$, however these are typically due to a strong softening of a phonon mode, indicating a possible dynamical instability of the structure.
Comparing to the anti-perovskites (mean $\lambda=0.36$), on average the Heuslers have a smaller value of $\lambda$. This can be explained by the presence of first-row non-metallic elements in the anti-perovskites (like C, N, or O), that have the tendency to form strong covalent bonds.

The distribution of $\olog$ is almost symmetric, a fact translated by the proximity of the mean and median values ($190$ and $192$~K, respectively), as well as the reduced value of the skewness ($0.20$).
The anti-perovskites show a comparable distribution, at least in qualitative terms. It is also quite symmetric but shifted towards higher values of the frequency range ($\omega_\text{log}$ mean of 234~K, median of 230~K). Given the presence of light elements in the stable anti-perovskites, and therefore higher overall phonon frequencies, this was to be expected.

Due to the interplay of factors involved in these two quantities ($\lambda$ and $\olog$), they present a loose inverse proportionality relation~\cite{10.1088/1361-648x/ac2864} (see Figure~S3 of the SM). This makes increasing the critical temperature a challenging job, since it implies the simultaneous maximization of both parameters. 
On average, the increased $\lambda$ of the anti-perovskites with respect to the Heuslers compensates the reduction in $\olog$, meaning that these materials lie in a higher Pareto front of the $(\lambda, \omega_\text{log})$ diagram, and therefore tend to have slightly higher critical temperatures.

Regardless, as seen in Figure~\ref{fig:lambda_omega_tc_histograms}, the majority of materials for both families lie in the region below $1$~\si{\kelvin}. Of the handful of outliers with temperatures above 2~K, we find 8 Heusler with $T_c^\text{Eliashberg}$ above 5~K. The highest of these is \ce{Nb2ReRu} for which $T_c^\text{Eliashberg}=9.9$~K ($T_c^\text{McMillan}=8.3$~K).

\subsection{Machine Learning}

The calculations of the previous section are limited to a very thin range of thermodynamic stability. However, meta-stable phases are known to be synthesizeable, making them of potential technological interest. Due to the number of materials in this energy range, we opted to accelerate the search by using machine learning models to screen more efficiently the composition space.

With the data from the electron-phonon calculations we trained two machine-learning models in an attempt to classify and understand the larger set of materials further from the hull. Our dataset, although large for superconductor standards, is small for the typical use case of machine learning methods. With this in mind we chose two models: Operon~\cite{burlacuOperonEfficientGenetic2020}, a framework for symbolic regression, and the model agnostic supervised local explanations (MAPLE)~\cite{plumbModelAgnosticSupervised2018}. Besides performing well for smaller datasets, these models have the added benefit of providing some interpretability from the learned model. In the following we discuss the results from each of the machine learning models.

\paragraph{Symbolic regression (Operon model)}
To build an analytical expression for the target properties as a function of the features via symbolic regression, we allowed for the following operators: multiplication, division, a constant, $\log$, $\sqrt{\,}$, \raisebox{1ex}{$\scriptstyle\wedge$}2 and $\exp$. The resulting formulas for each set are presented in Tables~SVI and SVII of the SM.

For $\omega_\text{log}$ the training of the model yielded the formula
\begin{equation}
    \omega_\text{log}=c_0+c_1\cdot\frac{\mathrm{Col}_X\, e^{-c_2\cdot\mathrm{Col}_X}}{V}
    \label{eq:operon_omega_log}
\end{equation}
in four out of the ten different runs, with all runs combined returning a mean absolute cross-validation error of 39.4~K.
This formula presents a rather simple dependence on just two quantities: the unit cell volume, $V$, and the periodic table column of element $X$, $\mathrm{Col}_X$.
The inverse proportionality on $V$ can be understood since large cell volumes usually translate into large atomic radii and therefore larger atomic masses (thus reducing phonon frequencies).
As for the column number, a dependence on this quantity is expected from the empirical Matthias' rules~\cite{matthiasChapterSuperconductivityPeriodic1957}, modulating a change in number of valence electrons uncorrelated to changes in atomic volumes and masses.
It is however curious to notice that Operon finds element \ce{X} to have a comparatively higher importance than the remaining elements.
Equation~\eqref{eq:operon_omega_log} has a saddle point for $\mathrm{Col}_X = 1/c_2$, which for $c_2\approx0.2$ (close to the value obtained from all Operon regressions giving this formula) gives a preference for group 5 for atom \ce{X}.

For the electron-phonon mass enhancement parameter, the formula
\begin{equation}
    \lambda=c_0+c_1\cdot V\sqrt{\mathrm{Col}_X\,\mathrm{DOS}(E_F)} 
    \label{eq:operon_lambda}
\end{equation}
appeared in six out of ten runs (and has a mean absolute cross-validation error of 0.114).
Independently of the training for $\omega_\text{log}$ a direct dependece on the same quantities was found for $\lambda$.
The increasing behaviour of $\lambda$ with the density of states is a feature of some models for $\lambda$ (not necessarily with the sub-linear scaling) and thus expected.
For example, assuming an Einstein solid the dependence $\lambda \propto \mathrm{DOS}(E_F)$ is obtained~\cite{engelsbergCoupledElectronPhononSystem1963}.

\begin{figure}
  \centering
  \includegraphics[width=\linewidth]{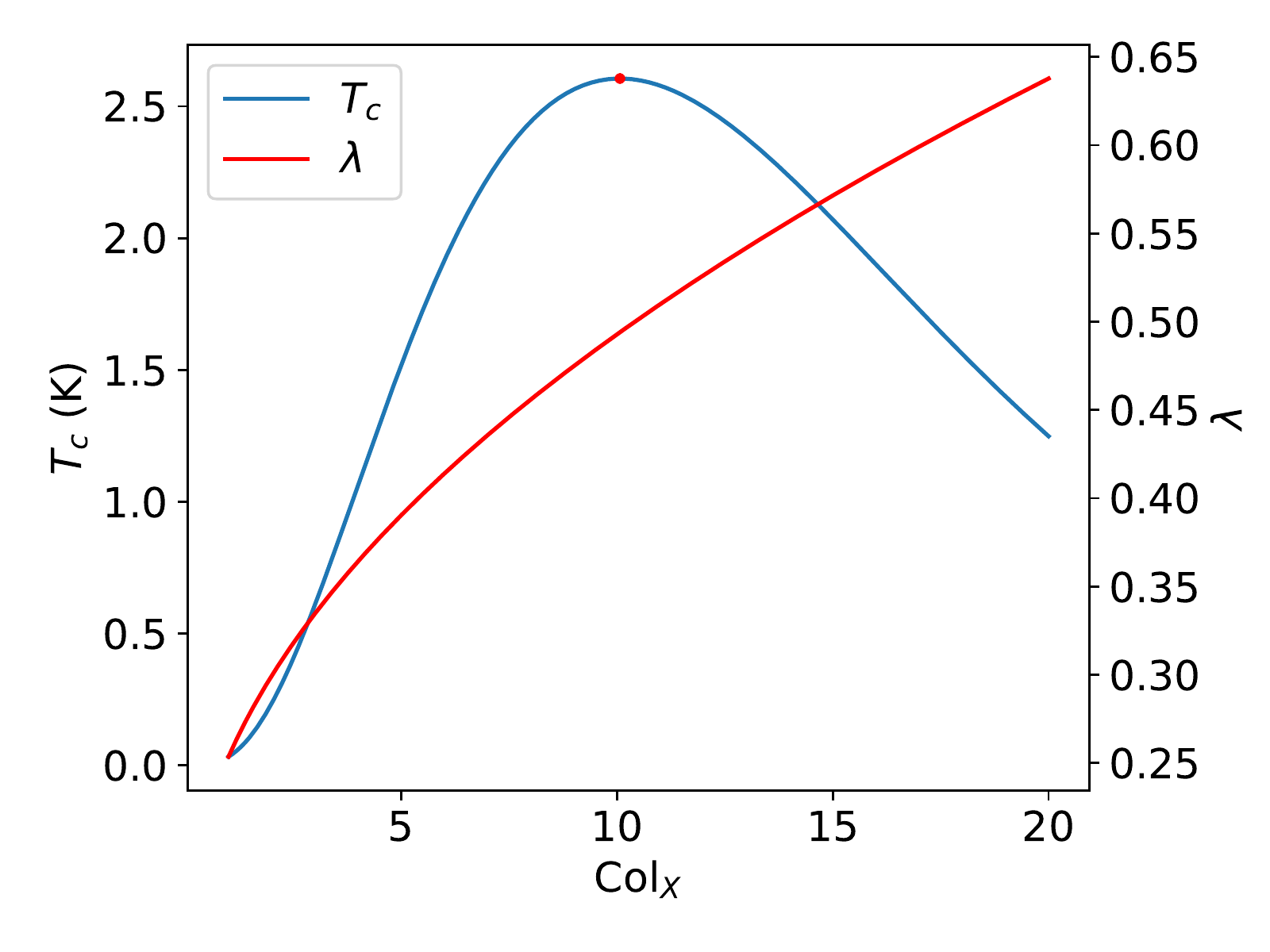}
  \caption{Plot of $T_c(\mathrm{Col}_X)$ and $\lambda(\mathrm{Col}_X)$ using the equations \eqref{eq:operon_lambda} and \eqref{eq:operon_omega_log}. The values for $\mathrm{DOS}(E_F)$ and the unit-cell volume were 2.66 (eV/states) and 85\,\AA$^3$, which are the mean values of all calculated systems. The red dot indicates the maximum value of \Tc.}
  \label{fig:Tc_vs_col}
\end{figure}

Assuming validity of these formulas for the Heusler family, we can study the evolution of the Allen-Dynes critical temperature formula as a function of the parameters found by Operon (see Figure~\ref{fig:Tc_vs_col}).
Taking $V$ and $\mathrm{DOS}(E_F)$ at their mean values, the maximum of $T_c$ is reached for group 10, that contains \ce{Ni}, \ce{Pd}, and \ce{Pt}.
This result is in line with previous experimental results, as most known Heusler superconductors with a high $T_c$ do contain \ce{Pd}.

\paragraph{Random forests (MAPLE model)}

MAPLE is a random forest based model capable of accurate predictions, while also providing some form of local interpretability.
Training the model to predict $\omega_{\log}$ yields an error of 34~K, better than the error of the Operon formulas. To express the equivalence of the Y and the Z atom we can double the data by exchanging their roles, which decreases the error to 27~K. The features with higher weight are the unit-cell volume and the column of the $X$ atom. Additionally, the MAPLE model shows a large weight for the total atomic mass of the compound. 

The models for $\lambda$ have a similar error as the Operon formulas, with a mean absolute cross-validation error of 0.11, regardless of the data doubling procedure. The most important features here are the density of states at the Fermi level, the unit-cell volume and, again, the column of the $X$ atom. Overall, the MAPLE analysis is in agreement with the feature importances returned by the Operon formulas.

With the trained MAPLE model we are in a position to widen our search of high-\Tc\ Heuslers to materials slightly further from the convex-hull. This is interesting as some of these compounds, with relatively small distances to the hull, might still be synthesizable experimentally. Furthermore, extending the number of materials studied gives us a better understanding of superconductivity of Heuslers and of the extreme values of \Tc\ that are attainable in this family. We therefore listed all compounds below 200~meV/atom from the convex hull that MAPLE predicted to have \Tc\ larger than 1~K. As seen from the distribution of the distance to the convex-hull available in Materials Project, this threshold encompasses the large majority (90\%) of the synthesized materials, making it a suitable limit for our search. These 749 materials were then validated by calculating the electron-phonon coupling using the 4x4x4 parameters (see \cref{sec:methods}).

It turns out that the large majority of materials exhibit imaginary frequencies, and are dynamically unstable. In fact, this was the case of 641 compounds out of the 749 compounds. A possible explanation to this fact is that MAPLE is predicting materials with very large values of $\lambda$ that are often dynamically unstable. As expected, the remaining 108 compounds have, on average, higher values of \Tc\ than the training set (see Table~SIV of the SM), with the appearance of a noticeable number of materials with temperatures above 5~K and even 10~K.

This shift is also accompanied by a slight qualitative change in the distribution of elements of high $T_c$ materials (see Figures~S5 and S6 of the SM). In the initial training set, a broad distribution of the chemical elements of the periodic table is observed, but with the materials predicted by machine-learning these concentrate around the earlier groups of the transition metals (Ti, V, Cr). The dependence of the mean \Tc\ is much stronger on the column than on the row, in agreement with the results from the machine learning models. This behaviour somewhat contrasts with the distribution of the anti-perovskites where the distribution is broader, favouring light elements like H, Be and N.

All information regarding these materials is readily available in the Tables presented in the Supplementary Material.

\subsection{Elastic constants and ductility}

\begin{figure}
  \centering
  \includegraphics[width=\linewidth]{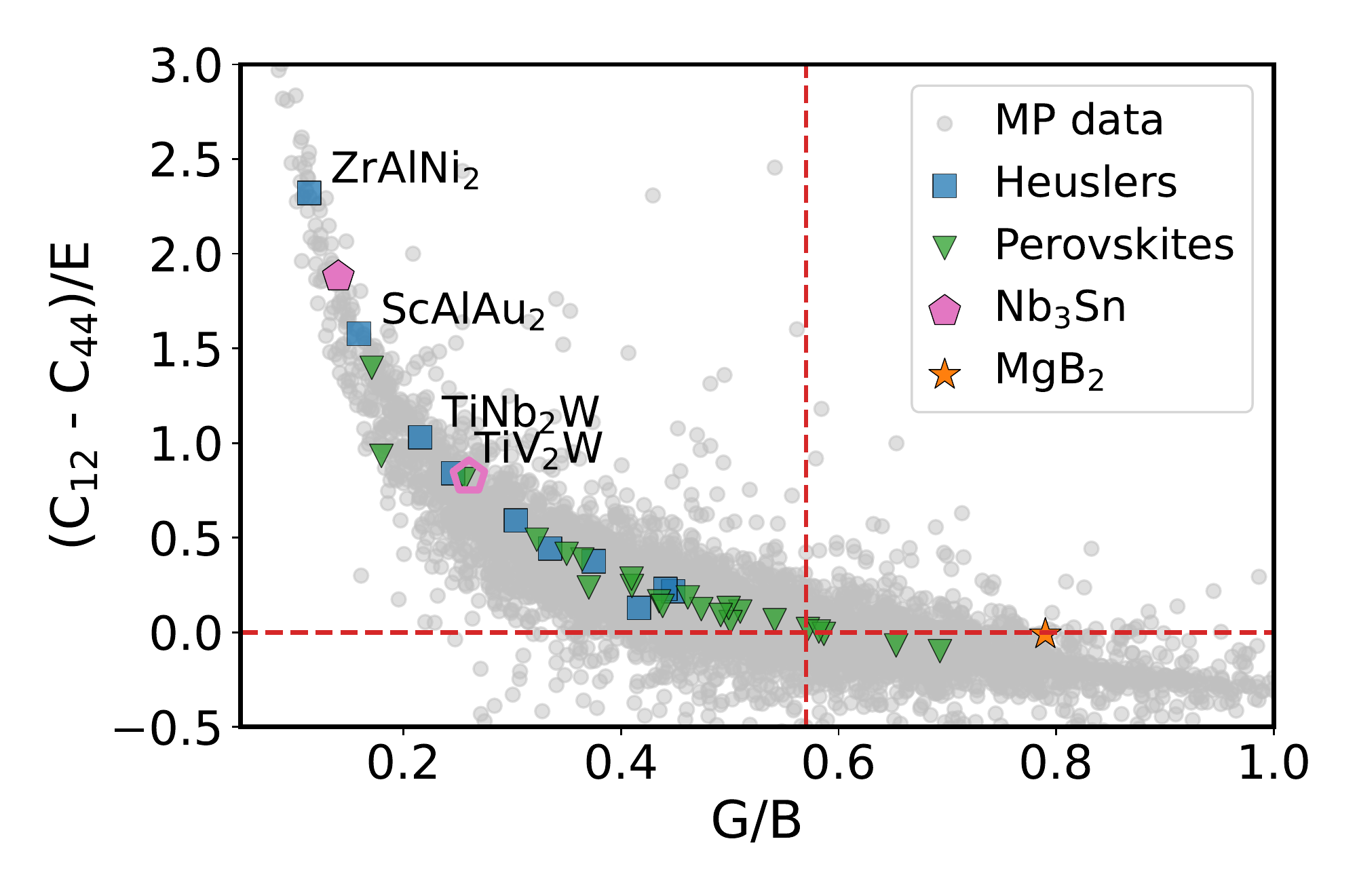}
  \caption{Dispersion plot of the ductile/brittle classification according to Petiffor's and Pugh's criteria. Green triangles and blue squares represent anti-perovskites and Heuskers, respectively, with $\Tc>5$~K and $E_\text{hull} < 50$~meV/atom. For reference, we also show the calculated values for \ce{MgB2} (orange star) and \ce{Nb3Sn} (pink pentagons).
  Grey squares represent entries from Materials Project~\cite{materialsproject}.
  Due to the considerable error in the theoretical determination of the elastic constants of \ce{Nb3Sn} we also show the experimental values from Ref.~\cite{carroll1965elastic} as an empty pentagon.}
  \label{fig:petiffor_vs_pugh}
\end{figure}

As the last step of our high-throughput analysis, we now turn to the discussion of the ductility of the superconducting Heuslers. For materials with \Tc\ above 5\,K we computed the stiffness tensor as described in \cref{sec:methods}. From these, we performed the ductility classification as described by Pugh's and Pettifor's criteria for both anti-perovskites and Heuslers. The results can be seen in \cref{fig:petiffor_vs_pugh} and the complete values are presented in the SM. Vickers hardness, $H_\text{V}$, was also estimated using the model from Ref.~\cite{Mazhnik2019}. Models of this type neglect important effects such as grain size, dislocations, etc. which have a notorious effect on hardness. Regardless, they seem to be accurate enough for the present purposes, as we only use them as a comparative proxy for the ductility of the material. Almost all of the materials are classified as ductile using the previous criteria. In addition, most of them also have large Poisson ratios, $\nu$, and low $H_\text{V}$, further hinting at their ductility.

With respect to anisotropy, we find a rather large range of values for the Zener ratio ($A$, see Sec.~\ref{sec:methods}), from around 0.2 to 4.3 (see SM). Some extreme cases occur for \ce{ZrAlNi2} (for which the extremely low $C_{44}=10$~GPa, translates simultaneously into low Zener ratio, shear modulus and Pugh's ratio~\cite{pughXCIIRelationsElastic1954}) and \ce{LiBe2Pt} (where all the constants have comparable magnitudes but the material still has a comparatively low shear modulus). 
As for the perovskites, considering the prevalence of elements that are associated with strong covalent (and therefore highly directional) bonds, these are expected to be brittler than the Heuslers, a fact corroborated by the ductility diagram \cref{fig:petiffor_vs_pugh}.

\begin{figure}
  \centering
  \includegraphics[width=\linewidth]{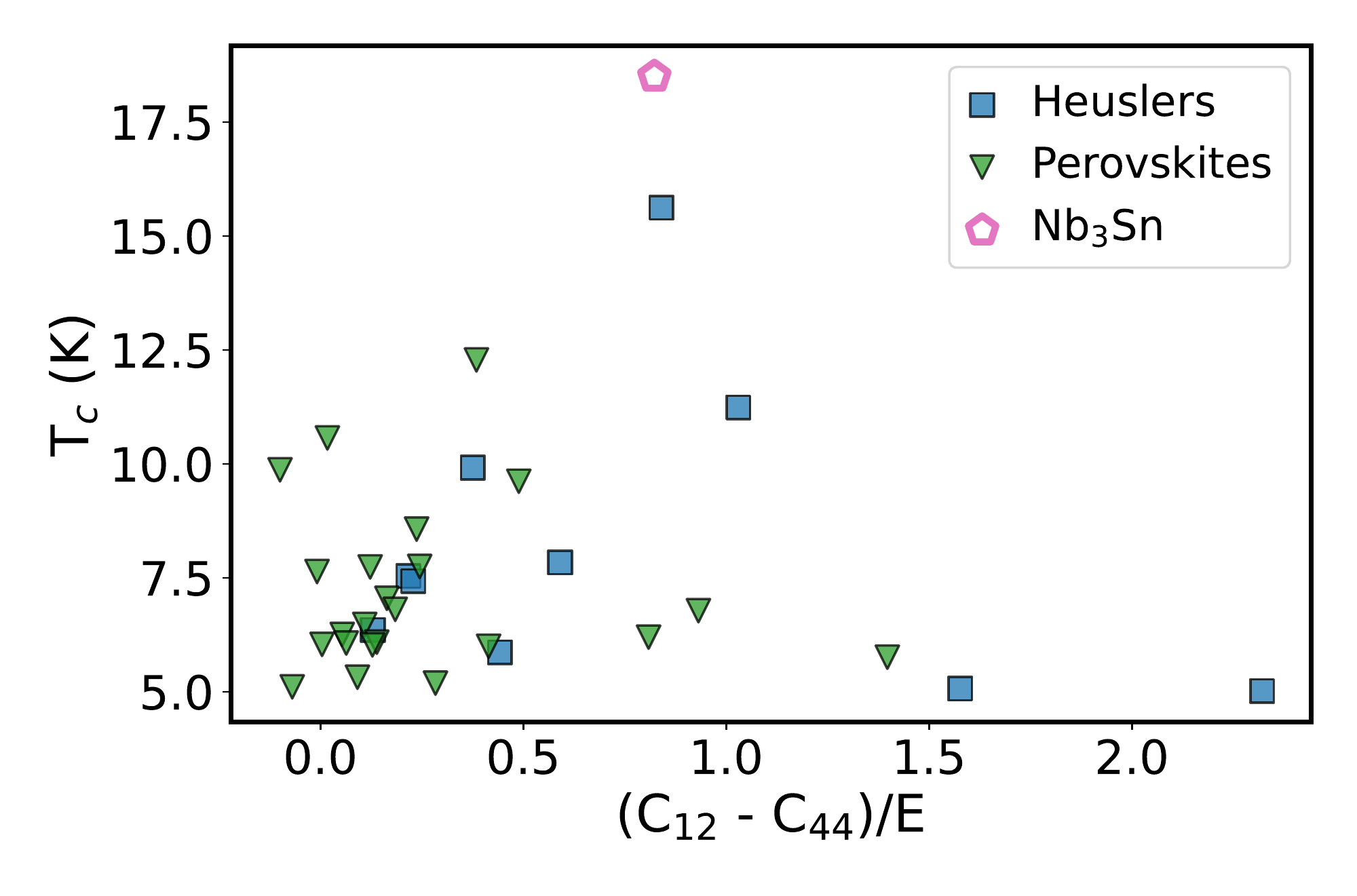}
  \caption{Dispersion plot of \Tc\ vs normalized Cauchy pressure, $(C_{12}-C_{44})/E$. Green triangles and blue squares represent anti-perovskites and Heuskers, respectively, with $\Tc>5$~K and $E_\text{hull} < 50$~meV/atom. For reference, we also show the experimental value for \ce{Nb3Sn}, shown as empty pentagon.}
  \label{fig:tc_vs_petiffor}
\end{figure}

\subsection{Individual entries}

Starting from the high-throughput calculations we selected the `best' materials within $\sim$50~meV/atom from the convex-hull for further analysis. These were chosen on the basis of the compromise between critical temperature and ductility. For these materials, more accurate calculations were performed with the tighter convergence parameters (see \cref{sec:methods}).
In the following, we discuss a couple of the selected materials, while the complete set of electronic and phononic bandstructures, along with other superconducting data, is given in the SM. 

\paragraph{\ce{V2TiMo}}

This compound has the merit badge of having the largest critical temperature of the present work, with $T_c^\text{Eliashberg}=19$~K ($T_c^\text{McMillan}=16$~K) and $E_\text{hull}=52$~meV/atom. However, we must notice that this compound includes vanadium that is known to lead to strong spin fluctuations~\cite{PhysRevB.40.8705} (not present in our approach), resulting in a noticeable decrease of \Tc.

\begin{figure}
  \centering
  \begin{tabular}{c}
    \includegraphics[width=0.97\linewidth]{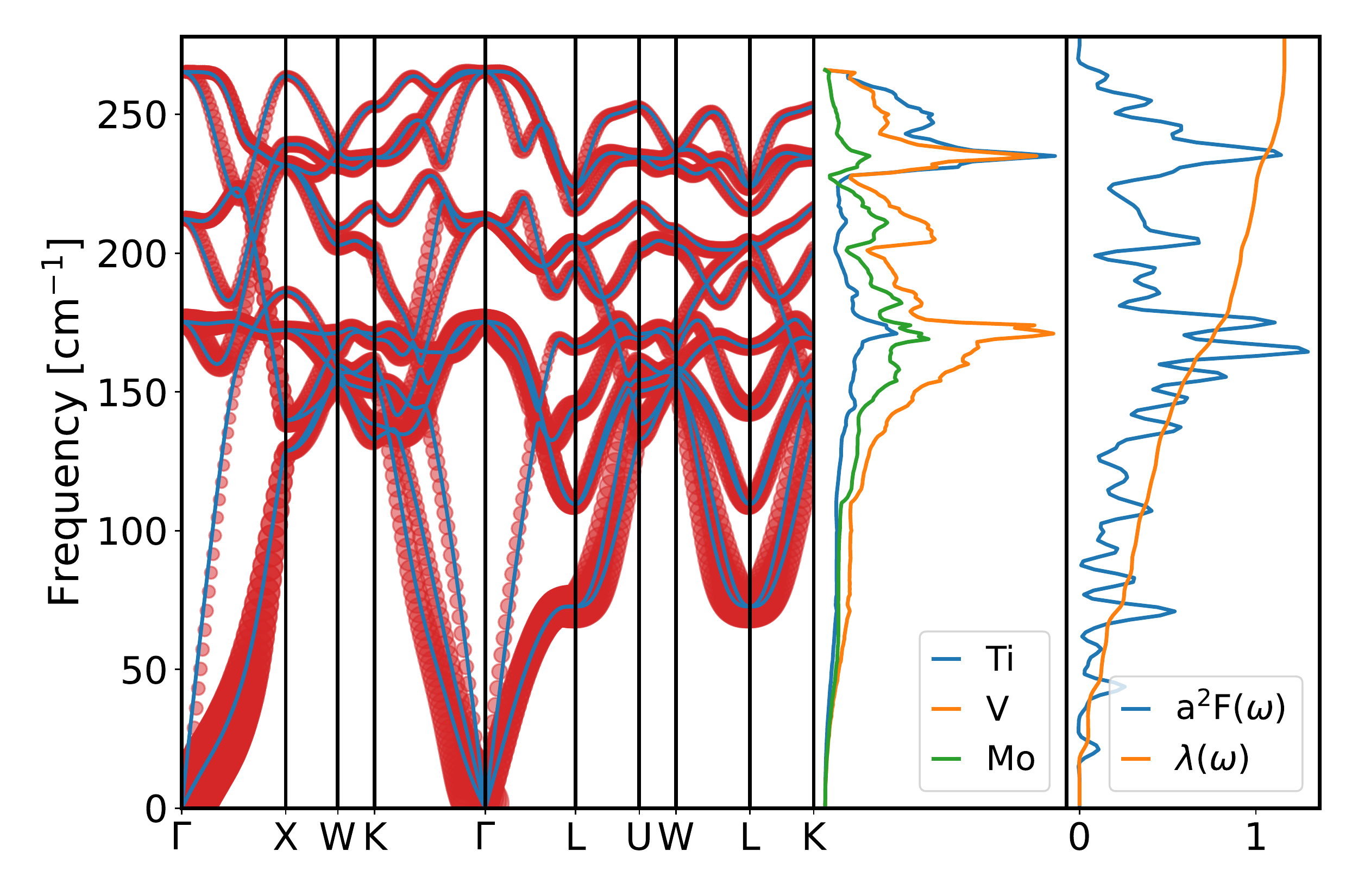} \\
    \includegraphics[width=0.97\linewidth]{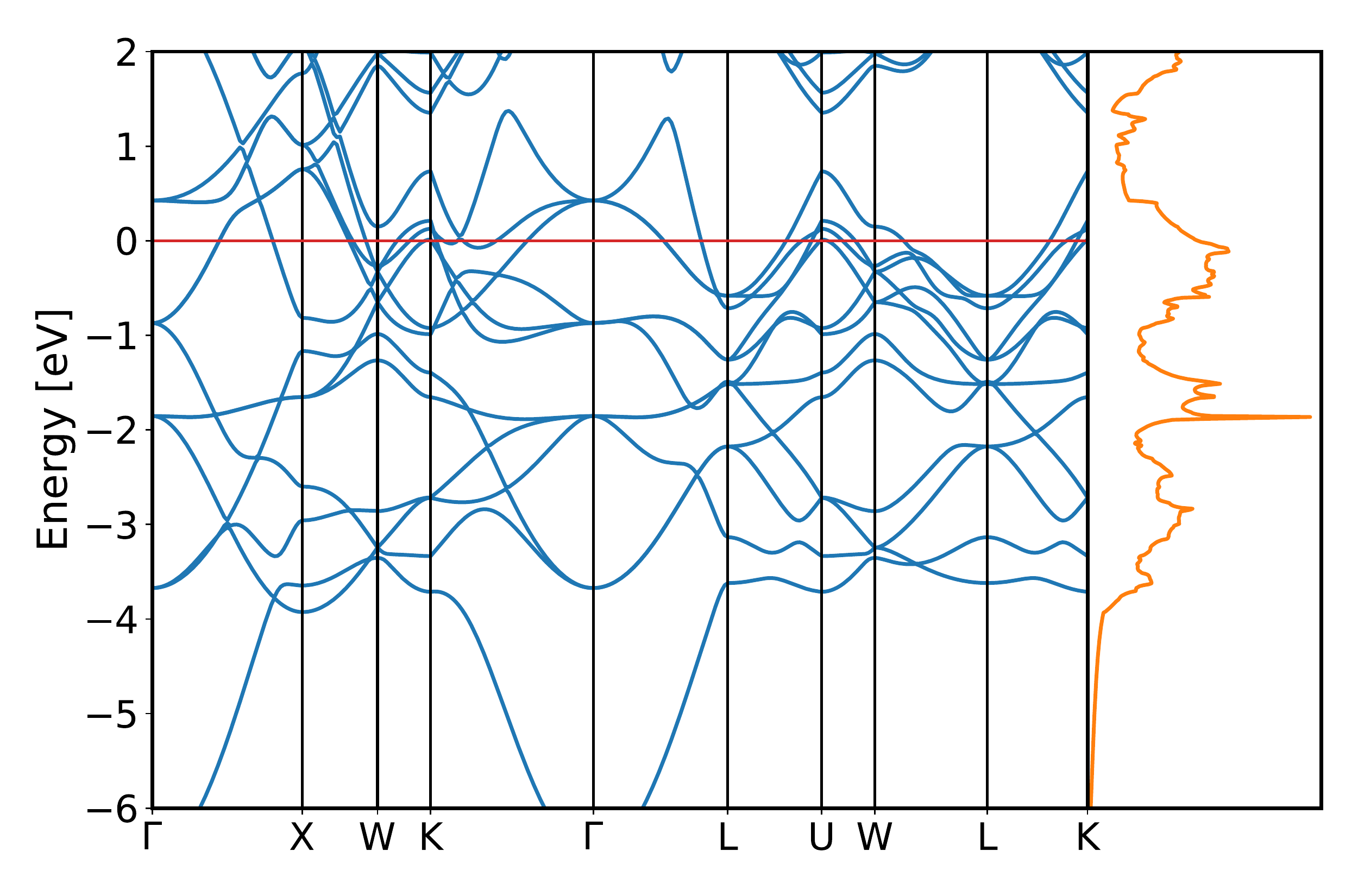}
  \end{tabular}
  \caption{Calculated phonon dispersion curves (along with atom-projected phonon density of states and Eliashberg spectral function)
  and electronic bandstructure (along with density of states) for \ce{V2TiMo}.
  Broadening in phonon bandstructure represents the magnitude of the electron-phonon coupling strength, $\lambda_{n\mathbf{q}}$.
  Origin of the energy in electronic plots has been shifted to the Fermi level.}
  \label{fig:V2TiMo_phonons_and_bands}
\end{figure}

Looking at the electronic band structure, shown in Figure~\ref{fig:V2TiMo_phonons_and_bands}, although the density of states near the Fermi level is large, it could be slightly increased via hole doping. In turn this is expected to translate in a small increase of $\lambda$ and therefore of $T_c$. This large density of states is due to a series of almost parabolic bands which interpenetrate close to the Fermi level, very similar in shape to that of several other materials here, for example \ce{Ti2NbRe}.

In the phonon bandstructure we find two notable peaks in the density of states, one close to 170~\si{\per\centi\meter} and the other one close to 230~\si{\per\centi\meter}. The former mostly results from V contributions while the later is due to both Ti and V in almost equal value.
Several phonon branches with strong electron-phonon coupling strength are observed, in particular associated with the two lowest acoustic and optical modes at $\Gamma$ as well as the lowest frequency modes at L. In total, this gives rise to an electron-phonon mass enhancement parameter of 1.2.
Due to the low density of states at low frequencies, the large $\lambda_{n\mathbf{q}}$ from these modes contribute only moderately to $\alpha^2F(\omega)$ when compared with the aforementioned two peaks. This results in $\omega_\text{log}$ reaching 189~K, which is not particularly large when compared to, for example, \ce{Be2CoNi}, but larger than that of  \ce{Ti2NbRe}.

Elastically, \ce{V2TiMo} lies comfortably in the ductile region ($G/B=0.2$ and $(C_{12}-C_{44})/E=1.3$). In spite of this, the small value of $C_{44}=21$~GPa translates into a very small Zener ratio of 0.3, i.e., a highly anisotropic elastic response under shear.

\paragraph{ \ce{Nb2TiW} }

With a lower $T_c^\text{Eliashberg}=11$~K ($T_c^\text{McMillan}=9$~K), \ce{Nb2TiMo} also lies lower in the ductility hyperbole ($G/B=0.2$ and $(C_{12}-C_{44})/E=1.0$) with the advantage of being energetically more stable, only 25~meV/atom from the hull.

Its electronic structure is remarkably similar to the one of \ce{V2TiMo}, showing essentially the same qualitative behaviour. The phonon bandstructure on the other hand differs, most notably in the fact that the peak in the density of states due to the contributions of W is shifted to lower frequencies from the peaks of the other elements.
Curiously this nonetheless leads to the same value of $\omega_\text{log}$ as \ce{V2TiMo} (187~K), but the lower electron-phonon coupling strength leads to the lower $\lambda$ of 0.8 and in turn to the aforementioned lower critical temperature.

\begin{figure}
  \centering
  \begin{tabular}{c}
    \includegraphics[width=0.97\linewidth]{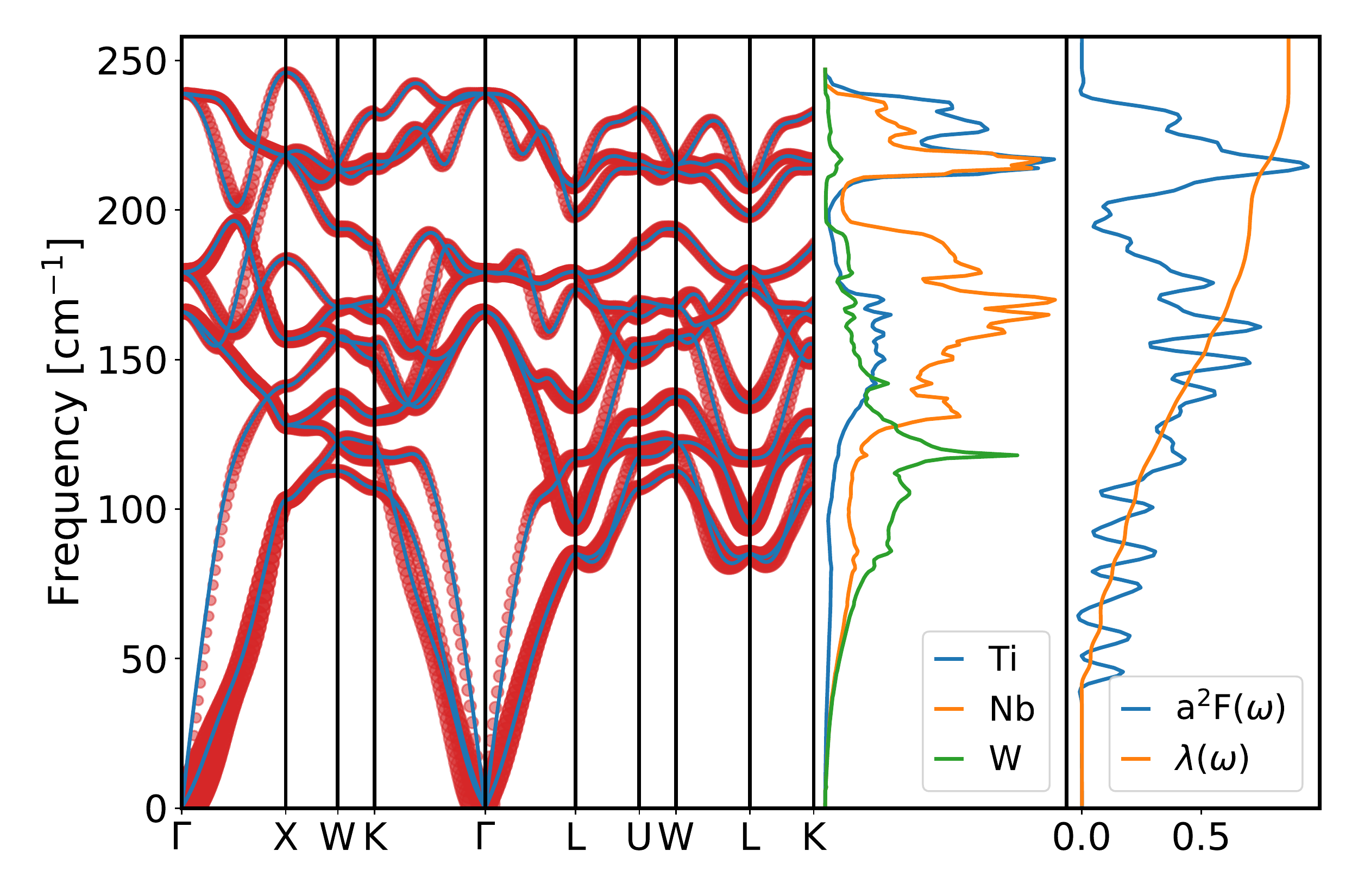} \\
    \includegraphics[width=0.97\linewidth]{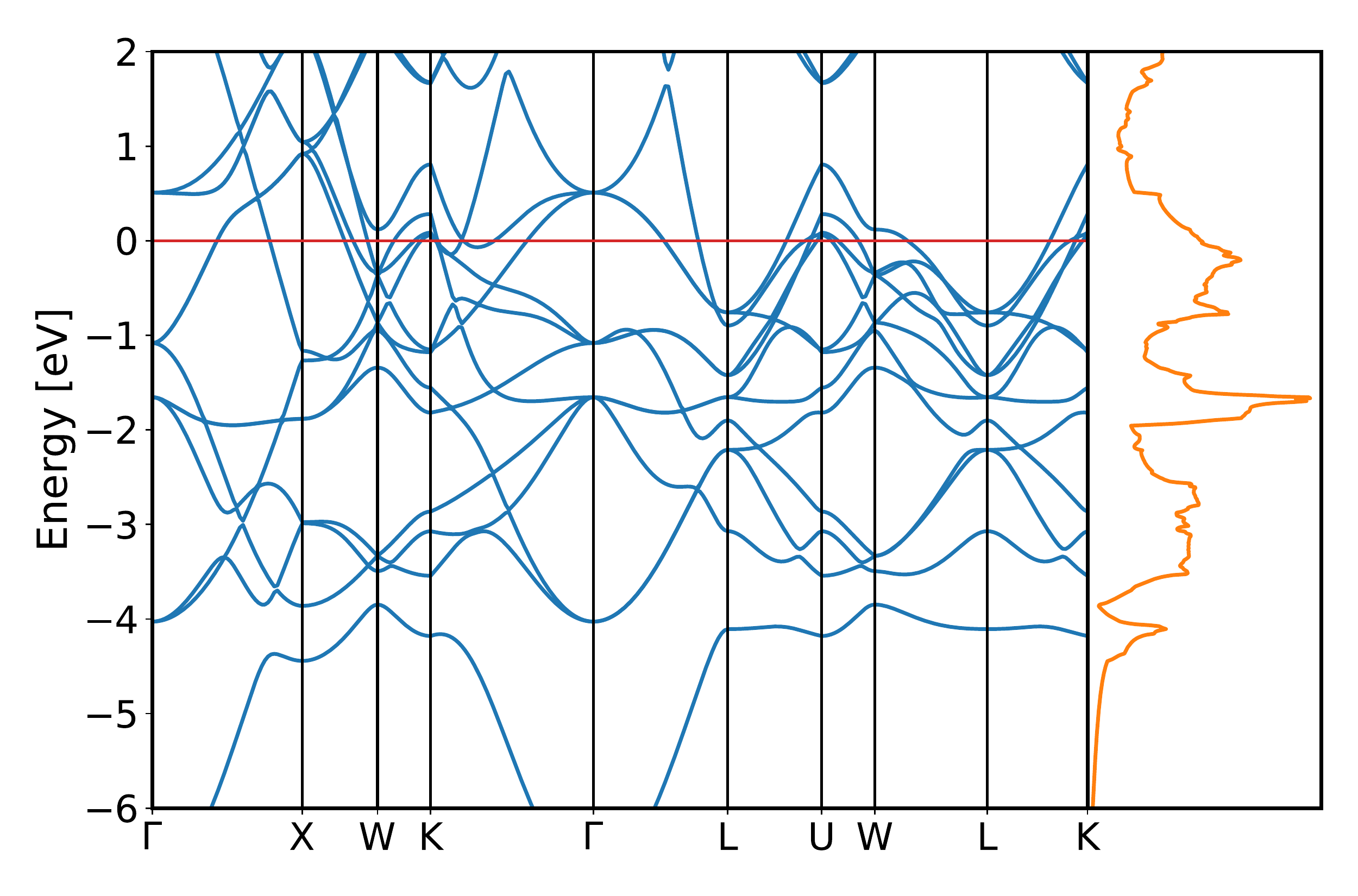}
  \end{tabular}
  \caption{Calculated phonon dispersion curves (along with atom-projected phonon density of states and Eliashberg spectral function)
  and electronic bandstructure (along with density of states) for \ce{Nb2TiW}.
  Broadening in phonon bandstructure represents the magnitude of the electron-phonon coupling strength, $\lambda_{n\mathbf{q}}$.
  Origin of the energy in electronic plots has been shifted to the Fermi level.}
  \label{fig:Nb2TiW_phonons_and_bands}
\end{figure}

\paragraph{ \ce{Nb2ReRu} }

Lower still in the ductility range lies \ce{Nb2ReRu}, specifically at $G/B=0.4$ and $(C_{12}-C_{44})/E=0.4$. In spite of its position, this material has the small advantage of presenting a much more isotropic elastic response than the previous entries ($A=1$).

\begin{figure}
  \centering
  \begin{tabular}{c}
    \includegraphics[width=0.97\linewidth]{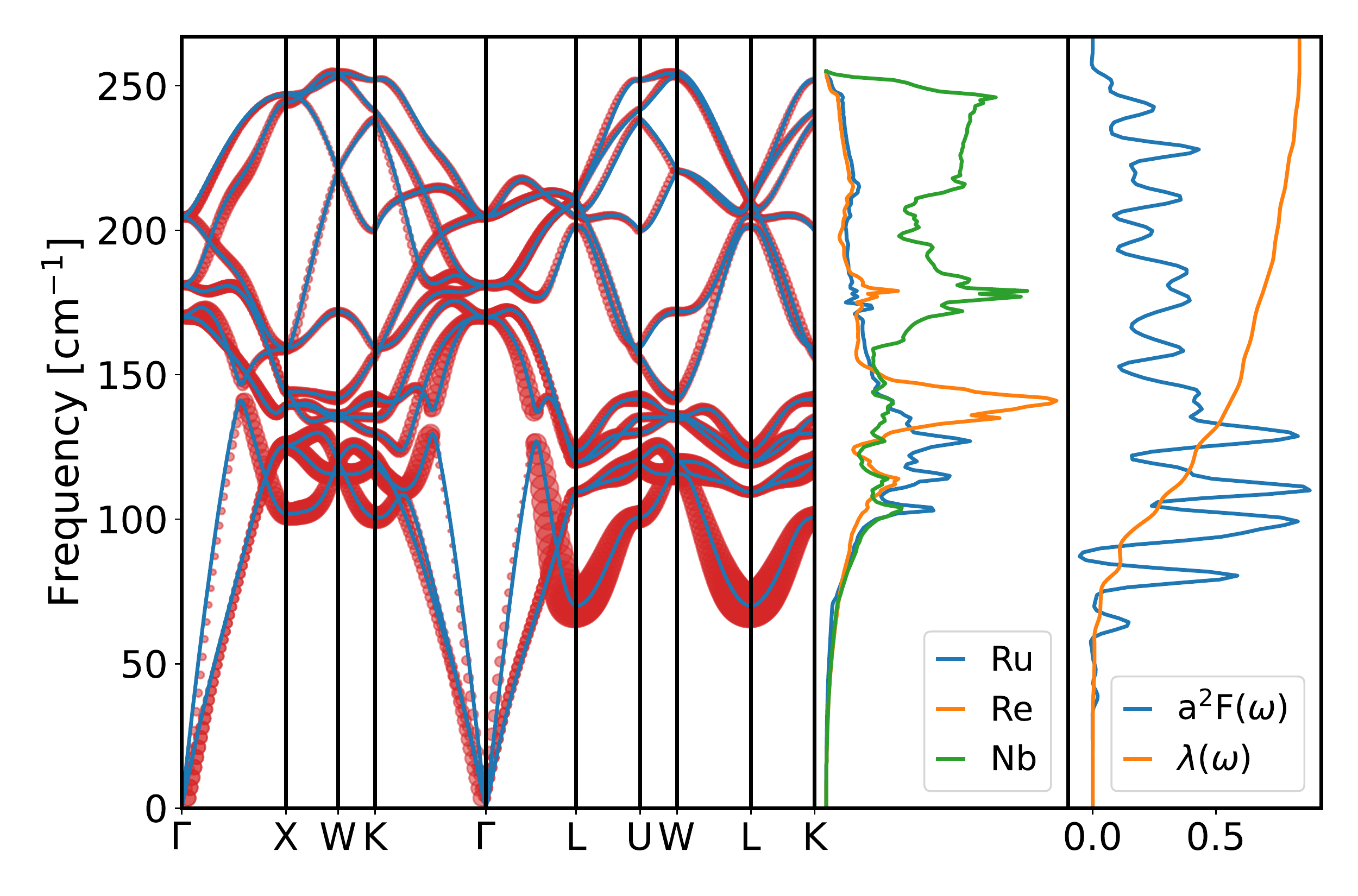} \\
    \includegraphics[width=0.97\linewidth]{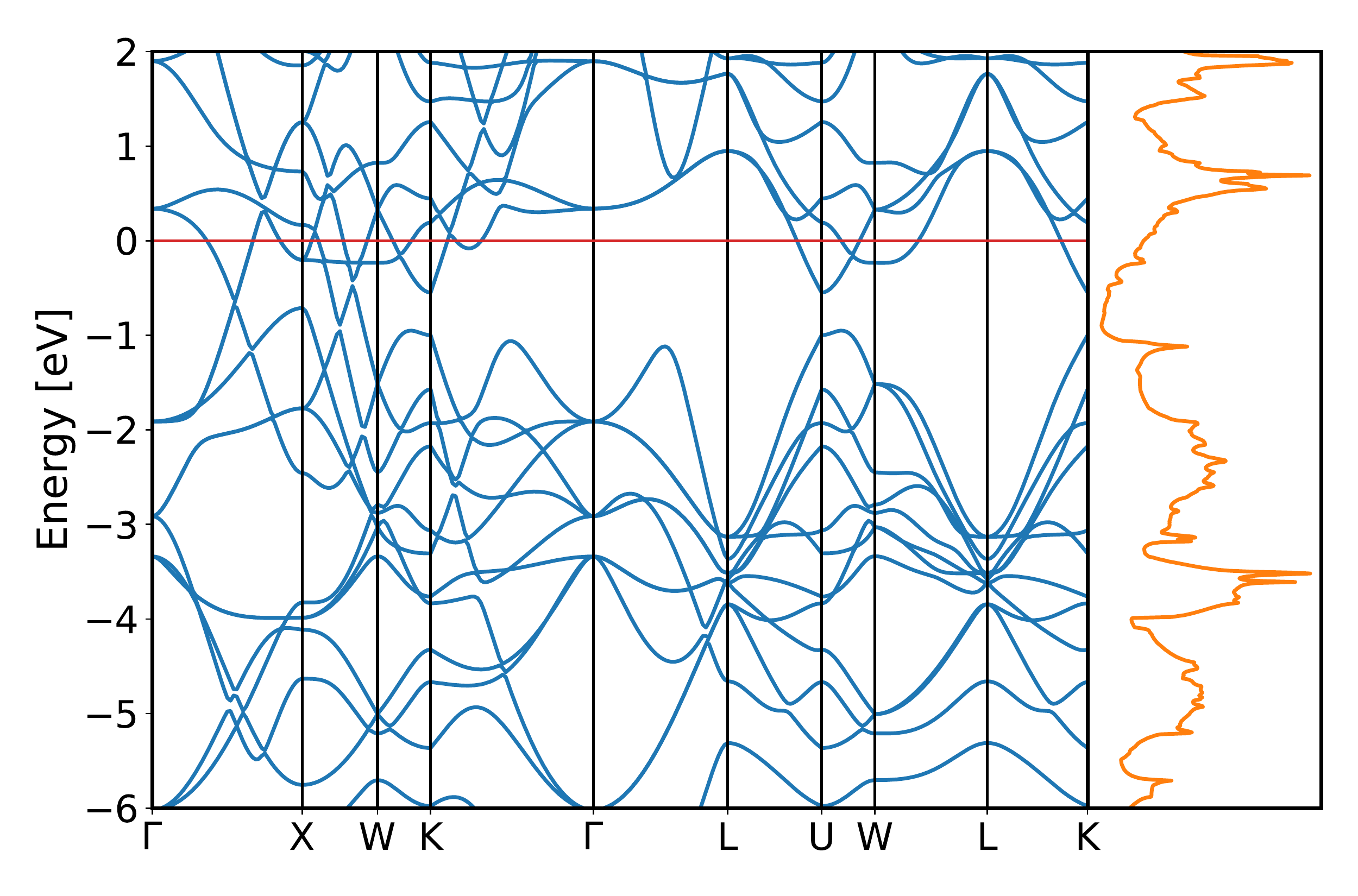}
  \end{tabular}
  \caption{Calculated phonon dispersion curves (along with atom-projected phonon density of states and Eliashberg spectral function)
  and electronic bandstructure (along with density of states) for \ce{Nb2ReRu}.
  Broadening in phonon bandstructure represents the magnitude of the electron-phonon coupling strength, $\lambda_{n\mathbf{q}}$.
  Origin of the energy in electronic plots has been shifted to the Fermi level.}
  \label{fig:Nb2ReRu_phonons_and_bands}
\end{figure}

Looking at the electronic band structure in Figure~\ref{fig:Nb2ReRu_phonons_and_bands} we see that the Fermi level lies in the middle of a `ramp' in the density of states. Immediately below it lies a sparsely populated energy range due to several band maxima in the neighbourhood of the L point which do not contribute to the Fermi surface. Above the Fermi level a complicated landscape appears which ultimately contributes to the large peak circa to 0.8~eV. Even if reaching this optimal position is unpractical, any level of electron doping would lead to an increase of the density of states at the Fermi level.

The atomic contributions to the phononic bandstructure are well differentiated, as seen from the corresponding density of states. Below 150~\si{\per\centi\meter} the largest contributions come from Ru and Re with Nb taking over above this point. The modes with the largest $\lambda_{n\mathbf{q}}$ are those close to the L direction, which in spite of the very small corresponding density-of-states lead to the largest contributions to $\alpha^2F(\omega)$. Because of this, the value of $\omega_\text{log}$ is low at 190~K and $\lambda$ at 0.8 is low compared to other top performing Heusler. As such, we reach the range of $T_c^\text{Eliashberg}$ of 10~K ($T_c^\text{McMillan}=8.3$~K).

\section{Conclusions}

In conclusion, we performed a thorough analysis of the superconducting properties of the full Heusler \ce{X2YZ} family. These results were then compared to anti-perovskites. Distributions of values of $\lambda$, $\olog$, and $\Tc$ have similar shapes in these two families, hinting at the universality of such distributions. Mean values, however, differ due to the different chemistry of both families. As expected from the Heusler family of intermetallics, we observed that the most favourable elements for superconductivity are transition metals, while the anti-perovskites favour the presence of lighter atoms (e.g. H, Be, N), which in turn correlates to higher average phonon frequencies.

As expected, the number of Heusler materials with critical temperature above 1~K is a small fraction of the total composition space. Regardless, 22 materials were found with critical temperature above 5~K and 8 of the found materials have critical temperatures above 10~K. This should be compared to the current record of $\Tc=4.7$~K in this family. These materials are furthermore expected to be ductile, making them promising for practical applications for the generation of high magnetic fields.

We also show the usefulness of machine learning models in the interpretation and exploration of the data. Approaches such as symbolic regression and random forests, which perform well for our small datasets, allow us to understand our results and to train predictive models.

Further work to include materials in different structural prototypes is now underway and will hopefully lead to more insight on the distribution and the universality of superconducting properties across compound space. Furthermore,  increasing the size of the superconducting datasets will lead to more general and accurate machine learning applications that have the potential to accelerate research in this field.

\section{Methods}
\label{sec:methods}

\subsection{Crystal structure}

The starting point for the present work is the crystal structure of Heuslers. This prototype, with chemical composition \ce{X2YZ}, crystalizes in the $Fm\bar{3}m$ space group (number 225) with the \ce{X} atoms located at the Wyckoff position $8c$ $(1/4, 1/4, 1/4)$, \ce{Y} at position $4a$ $(0,0,0)$ and \ce{Z} at $4b$ $(1/2, 1/2, 1/2)$~\cite{aroyoBilbaoCrystallographicServer2006, aroyoBilbaoCrystallographicServer2006a}. To denote the Heusler compounds we use the notation \ce{X2YZ} where the Y is the transition metal and the Z is the main group metal~\cite{Bai2012_1230006}.

\subsection{Ground-state}

Distances to the convex hull within the Perdew-Burke-Ernzerhof (PBE)~\cite{pbe,pbe_erratum} approximation were recalculated with the convex hull of Ref.~\onlinecite{CGAT}. We note that this hull is considerably larger than the one of the Materials Project~\cite{materialsproject}. Furthermore, due to the recent updates to the hull~\cite{CGAT}, some of the compounds that were thermodynamic stable at the beginning of the present work now have positive distances to the convex hull. For completeness, we also present distances to the hull calculated with the PBE for solids~\cite{pbesol,pbesol_erratum} and SCAN functionals~\cite{SCAN} in the SM, following the approach and the convex hull of Ref.~\cite{dataset}

\subsection{Electron-phonon}

\begin{table}[htb]
  \centering
  \begin{tabular}{rcccc}
  material      & 3x3x3 & 4x4x4 & 6x6x6 & 8x8x8 \\ \hline
  \ce{AlScAu2}  & 3.51  & 4.06  &  4.16 &  4.15 \\
  \ce{BeSiOs2}  & 1.77  & 2.69  &  1.82 &  1.82 \\   
  \ce{LiPtBe2}  & 3.53  & 6.48  &  4.33 &  4.31 \\
  \ce{RuReNb2}  & 9.51  & 8.69  &  9.18 &  9.12 \\
  \end{tabular}
  \caption{Convergence of the transition temperature (obtained with McMillan's formula, in K) with respect to the calculation parameters. The label corresponds to the number of $q$-points used.}
  \label{tab:convergence}
\end{table}

We employed essentially the same workflow and convergence criteria as in our previous work on inverted perovskites~\cite{oursperovskites}. In this way we could directly compare the two families, as well as accumulate a consistent dataset of superconducting calculations. In short, we performed calculations using \textsc{quantum espresso} version 6.8 using pseudopotentials from the \textsc{pseudodojo} project~\cite{vanSetten2018pseudodojo}, specifically the stringent norm-conserving set. We used the high plane-wave cutoff energy as specified in \textsc{pseudodojo}. Self-consistent ground-state calculations were performed with a Gaussian smearing of 0.02~Ry until the energy was converged to $10^{-9}$~Ry. Geometry optimization was stopped when the forces on the atoms were smaller than $10^{-4}$~Ry/bohr, stresses smaller than 0.05~kbar, and when the difference of energy was smaller than $10^{-5}$~Ry. The threshold for self-consistency in the phonon calculations was set to $10^{-14}$. For the  calculation of the superconducting properties we used the Perdew-Wang~\cite{PerdewWangLDA} local-density approximation. In contrast to Ref.~\cite{oursperovskites} we employed the double $\delta$-integration to obtain the Eliashberg function in order to improve the accuracy of the calculations. To select the $k$- and $q$- point meshes, we performed convergence tests for 4 materials (see Table~\ref{tab:convergence}). The meaning of the columns is ``3x3x3'': coarse $k$-point grid 6x6x6, fine $k$-point grid 18x18x18, $q$-point grid 3x3x3; ``4x4x4'': coarse $k$-point grid 8x8x8, fine $k$-point grid 24x24x24, $q$-point grid 4x4x4; ``6x6x6'': coarse $k$-point grid 12x12x12, fine $k$-point grid 36x36x36, $q$-point grid 6x6x6; ``8x8x8'': coarse $k$-point grid 16x16x16, fine $k$-point grid 48x48x48, $q$-point grid 8x8x8. We can clearly see that the 8x8x8 are perfectly converged, while with a 4x4x4 $q$-grid one can already obtain a good approximation to $T_\text{c}$. Actually already with a 3x3x3 results are meaningful. As such, we decided to use the 4x4x4 for the high-throughput search, and the 6x6x6 for the systems we discuss in more detail.

\subsection{Superconductivity}

The values of $\lambda$, $\omega_2$ and $\omega_\text{log}$ were used to calculate the superconducting transition temperature using the McMillan formula~\cite{Mcmillan1968tc,Dynes1972}
\begin{equation}
    T_\text{c}^\text{McMillan} = \frac{w_\text{log}}{1.20} \exp\left[-1.04\frac{1 + \lambda}{\lambda - \mu^*(1 + 0.62\lambda)}\right]
    \,,
\end{equation}
and the Allen-Dynes modification~\cite{Allen1975} to it:
\begin{equation}
    T_\text{c}^\text{AD} = f_1 f_2 T_\text{c}^\text{McMillan}
    \,,
\end{equation}
where the corrections factor are
\begin{subequations}\begin{align}
    f_1 & = \left\{1 + \left[\frac{\lambda}{2.46(1 + 3.8*\mu^*)}\right]^{3/2}\right\}^{1/3} \, , \\
    f_2 & = 1 + \frac{\lambda^2 (\omega_2/\omega_\text{log} - 1)}
    {\lambda^2 + \left[1.82(1 + 6.3\mu^*) \omega_2/\omega_\text{log}\right]^2} \, .
\end{align}\end{subequations}

We took arbitrarily the value of $\mu^*=0.10$ for all materials studied. We note that this procedure is well defined for McMillan's and Allen-Dynes' formulas, but not for the Eliashberg equations. Indeed, these depend on an extra parameter, the cutoff of the Coulomb interaction, and for which we took the (rather arbitrary) value of 0.5~eV.

\subsection{Elastic constants}
\label{sec:methods_elastic_constants}

In order to study the elastic response of the materials under study, we computed the stiffness tensors via finite differences. Specifically, the elastic constants were obtained by computing the stress of a sufficient set of deformed structures and fitting the resulting values via Hooke's law. This entire procedure was done as implemented in the \textsc{thermo\_pw} package~\cite{thermo_pw}. The underlying calculations were done with \textsc{quantum espresso} version 6.8 and the Perdew-Wang local-density approximation to the exchange-correlation potential~\cite{PerdewWangLDA} . We resorted to the corresponding \textsc{stringent} set of norm-conserving pseudopotentials from \textsc{pseudodojo}~\cite{vanSetten2018pseudodojo}, from which the largest recommended energy cut-offs were chosen. To assure convergence with respect to the $k$-point sampling~\cite{deJong2015}, a constant value of 12000 $k$-points per reciprocal atom~\cite{Ong2013pymatgen} was used for all materials. 

Due to the directional dependence of the elastic responses, an averaging method is recommended for large-scale analysis. Following Materials Project we resort to the Voigt-Reuss-Hill~\cite{Reuss1929, Voigt1966, Hill1952} average, i.e., the simple average of the higher and lower limits of the response for polycrystalline materials.
Anisotropy is another relevant quantity to consider, due to the correlation of the different spatial response with the appearance of failures such as cracks on crystals under stress. Since in this work we are dealing with cubic materials, the Zener ratio,
\begin{equation}
    A = \frac{2C_{44}}{C_{11}-C_{12}} \, ,
\end{equation}
is a sufficient quantity to study this. The ratio of the Voigt and Reuss shear moduli also provides a measure of this effect. Both of these quantities are 1 for an isotropic solid, thus giving a simple measure of the anisotropy. Vickers hardness, $H_\text{V}$ was also estimated using the model from Ref.~\cite{Mazhnik2019}.

Because of its importance in all areas depending on metallurgy, ductility is well understood at the macroscopic level. However, from an atomistic point-of-view describing ductility is not trivial. Several qualitative models exist based on the type of bonding but a more attractive approach are models based on elastic properties, readily available from \textit{ab-initio} methods.
According to Pugh~\cite{pughXCIIRelationsElastic1954}, the ratio $G/B$ (where $G$ is the shear modulus and $B$ the bulk modulus) gives a measure of the brittleness of the material; the smaller the ratio the more ductile the material. Several works propose different values for the `critical' ratio defining the onset of brittleness depending on the class of materials under study. The typical value seen in the literature is $0.57$ but due to the empirical nature of the parameter, this value might not be general. A recent work~\cite{wuJoAP2019} suggests the value $0.44$ (Christensen’s criterion~\cite{Christensen2013}) as a more realistic threshold for heuslers.
Pettifor's criteria~\cite{pettiforTheoreticalPredictionsStructure1992} measures the `directionality' of the bonds via the value of the Cauchy pressure, $C_{12}-C_{44}$, commonly normalized to the Young's modulus, $(C_{12}-C_{44})/E$. Negative values indicate directional bonding associated with brittle behaviour.
These two criteria can then be used to define a region of interest for ductility.

\subsection{Machine Learning}

For each \ce{X2YZ} entry we use as input feature of the models a mixture of structural and atomic properties. For the former we resorted to the volume of the unit cell ($V$), density of states at the Fermi level ($\mathrm{DOS}(E_F)$) and total atomic weight of the compound ($M_{\text{tot}}$) while for the later we used each atom's charge ($Q_\Lambda$, where $\Lambda\in\{X,\,Y,\,Z\}$), row and column in the periodic table ($\mathrm{Row}_\Lambda$ and $\mathrm{Col}_\Lambda$), electronegativity ($\chi_\Lambda$), relative atomic masses ($m_\Lambda=M_\Lambda/M_{\text{tot}}$) and covalent radius ($R_\Lambda$). As the atoms \ce{Y} and \ce{Z} are equivalent they were sorted by electronegativity, such that $\chi_Y<\chi_Z$.

Instead of training directly for \Tc\ we targeted $\lambda$ and $\olog$ independently in an attempt to minimize errors. Due to the relatively small size of the data set we used ten-fold cross-validation, randomly splitting the data into a training and validation (in a 80:20 ratio) set. The models were trained in each of the ten independent sets, with the mean of the errors on the corresponding validation sets being the cross-validation error. The several resulting formulas obtained for $\olog$ and $\lambda$ for the different training sets are presented in the SM.

For the symbolic regression with Operon we allowed for the following operators: multiplication, division, a constant, $\log$, $\sqrt{\,}$, \raisebox{1ex}{$\scriptstyle\wedge$}2 and $\exp$. In the following we will only mention parameters, which were changed, i.e. any unnamed parameters were left at the default value. We used 100 local iterations, a population size of 2000, 5000 generations. The expression tree was limited to a maximum depth of 10 and a maximum length of 6 and we optimised the mean square error.

MAPLE was used with random forests, 300 estimators, 50\% maximum feature participation, a minimum of 10 samples per leaf and a regularization of 0.001.

\section{Data availability statement}
All data used in or resulting from this work is available in the manuscript and the Supplementary Material.

\section{Acknowledgements}

T.F.T.C and P.B. acknowledge financial support from FCT - Fundação para a Ciência e Tecnologia, Portugal (projects UIDB/04564/2020 and 2022.09975.PTDC and contract 2020.04225.CEECIND) and the Laboratory for Advanced Computing at University of Coimbra for providing HPC resources that have contributed to the research results reported within this paper.

\section{Author  Contributions}

T.F.T.C., P.B., and M.A.L.M. performed the \emph{ab initio} calculations. N.H. and J.S. trained the machine learning models. A.S. developed the Eliashberg solver. All authors contributed to designing the research, interpreting the results and writing of the manuscript.

\section{Competing  Interests}

The authors declare that they have no competing interests.

\bibliographystyle{naturemag}
\bibliography{references.bib}

\end{document}